\documentclass[aps,
prb,
twocolumn,
showpacs,
superscriptaddress,
groupedaddress,
floatfix, 10pt]{revtex4-1}  

\usepackage{graphicx}
\usepackage{bm}
\usepackage{amssymb}
\usepackage{commath}
\usepackage{hyperref}
\usepackage{physics}
\usepackage[capitalize]{cleveref}

\begin{document}
\title{Manifestations of spin-orbit coupling in a cuprate superconductor}
   
\author{Zachary M. Raines}
\email{raineszm@gmail.com}
\author{Andrew A. Allocca}
\author{Victor M. Galitski}
\affiliation{Joint Quantum Institute and Condensed Matter Theory Center, Department of Physics, University of Maryland, College Park, Maryland 20742-4111, U.S.A.} 
\begin{abstract}
Exciting new work on Bi2212 shows the presence of non-trivial spin-orbit coupling effects as seen in spin resolved ARPES data [Gotlieb {\emph{et al.}}, \emph{Science}, \textbf{362}, 1271-1275 (2018)].
Motivated by these observations we consider how the picture of spin-orbit coupling through local inversion symmetry breaking might be observed in cuprate superconductors.
Furthermore, we examine two spin-orbit driven effects, the spin-Hall effect and the Edelstein effect, focusing on the details of their realizations within both the normal and superconducting states. 
\end{abstract}
\maketitle

\section{Introduction}

Since their discovery three decades ago, the cuprate family of superconductors has been a focus of intense research interest.\cite{bednorz_possible_1986}
To this day they maintain the highest superconducting transition temperature at ambient pressure.\cite{schilling_superconductivity_1993}
Despite many years of active investigation the cuprates continue to generate new discoveries and new puzzles.
Discussion continues on phenomena ranging from the exact nature of the pairing mechanism, to the origin of the pseudogap,\cite{lee_amperean_2014,badoux_change_2016,chatterjee_fractionalized_2016} and the various charge-ordered states now being observed.\cite{chang_direct_2012,ghiringhelli_long-range_2012,fujita_direct_2014,comin_symmetry_2015}

While there has been some work on the consequences of spin-orbit coupling in the cuprates, it is generally believed that such effects are weak\cite{edelstein_spin_1990,koshibae_electronic_1993,edelstein_magnetoelectric_1995,harrison_nodal_2015,wu_spin-orbit_2005} and they are often ignored. 
However, recent spin-resolved ARPES measurements have shown striking evidence of spin textures in the Brillouin zone.\cite{Gotlieb2018}
In particular, the observed behavior can be explained by a model of spin-orbit coupling which is opposite on the two layers of the BSCCO unit cell.
Such a model preserves the inversion symmetry of system but can still host non-trivial effects arising from the spin-orbit coupling.
There is precedent for superconductors with such a staggered noncentrosymmetry,\cite{sigrist_superconductors_2014} but the consequences for the cuprate superconductors have not yet been investigated.

It is well established both theoretically \cite{dyakonov_current_1971, hirsch_spin_1999, murakami_dissipationless_2003, sinova_universal_2004, belkov_magneto_2008} and experimentally\cite{kato_observation_2004, wunderlich_experimental_2005, zhao_coherence_2006} that systems with spin-orbit coupling may display novel transport properties linking spin and charge degrees of freedom.
These are typically called spintronic effects, and provide a potential means to manipulate spins with charges and vice versa.\cite{hoffmann_spin_2013, sinova_spin_2015}
One of the most commonly considered of these effects is the spin-Hall effect, in which a net spin polarization accumulates at the boundaries of a sample in response to an electric current.
As in the case of the Hall effect, the spin-Hall effect can also be related to a transverse current associated with the accumulated quantity, although in the case of a superconductor the relation between the two viewpoints is more subtle.
Another notable phenomenon is the Edelstein or inverse spin-galvanic effect, which relates a spin polarization throughout the bulk of a sample to the flow of a charge current.\cite{edelstein_spin_1990}
In light of the observations of spin-orbit coupling in cuprate superconductors, the question naturally arises as to how such effects manifest in these materials, particularly within the superconducting phase. 

The structure of this paper is as follows.
In Section~\ref{sec:model} we introduce the model of spin-orbit coupling in BSCCO and discuss some of its properties.
In Section~\ref{sec:bdg} we review the theory of superconductivity in spin-orbit coupled materials and construct a general Bogoliubov-de Gennes Hamiltonian describing $d$-wave superconductivity in this model.
In Sections~\ref{sec:she} and \ref{sec:ee} we then use this model to calculate spintronic effects in both the normal and superconducting states.
In Section~\ref{sec:end} we review and discuss our results.

\section{Model}
\label{sec:model}

In this work we use a model which was introduced to explain the experimentally observed spin-texture in Bi2212 under spin-resolved ARPES.\footnote{See the Supplemental material for Ref.~\onlinecite{Gotlieb2018}}
The model is a tight-binding description of the copper sites in a bilayer of CuO$_2$ planes.
It is given by
\begin{equation}
    H_0 = \sum_{\mathbf k} c^\dagger_{\mathbf k} \left(\xi(\mathbf k) + t_\perp(\mathbf k) \tau_x
    + \bm{\lambda}(\mathbf k) \cdot \bm{\sigma} \tau_z\right)c_{\mathbf{k}},
    \label{eq:H0}
\end{equation}
where $\xi(\mathbf k)$ includes hopping terms up to third-nearest neighbor, leading to a hole-like Fermi surface, $\bm{\lambda}(\mathbf k) = \lambda (\sin k_y, -\sin k_x, 0)$ is the spin-orbit coupling vector, and $t_\perp(\mathbf k) = t_\perp (\cos k_x - \cos k_y)^2$ is the interlayer hopping term.\cite{xiang_universal_2000,markiewicz_one-band_2005}
Here $\tau$ and $\sigma$ represent two sets of Pauli matrices, with the $\tau$ matrices operating in the layer space and the $\sigma$ matrices operating in spin space, and $c_\mathbf{k}$ is the 4-component vector of electron annihilation operators in the spin and layer spaces.
The form of the spin-orbit coupling can be ascribed to local inversion symmetry breaking between the layers; field effects in between the layers lead to inversion symmetry breaking with opposite sense in the top and bottom layer, such that the system as a whole retains inversion symmetry.

\begin{figure}
    \centering
    \includegraphics[width=0.75\linewidth]{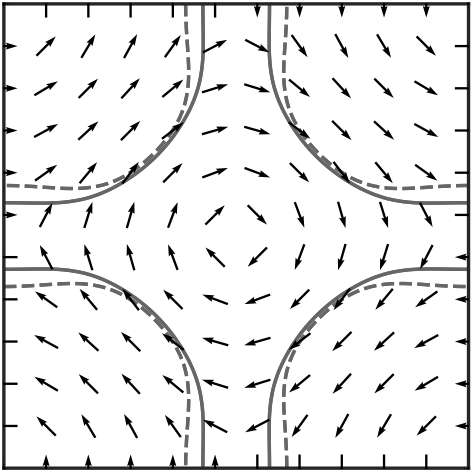}
    \caption{The Fermi surfaces for the two bands with spin aligned along the spin-orbit coupling vector $\bm{\lambda}(\mathbf k)$.
    For the opposite helicity the two bands are exchanged.
    The spin-texture for the band parallel to the spin-orbit vector is shown by the arrows. \label{fig:fermi_surface}}
\end{figure}

This system contains two Kramers degenerate bands with energies
\begin{equation}
    \epsilon_b(\mathbf k) = \xi(\mathbf k) + b A(\mathbf k),
\end{equation}
where $b=\pm$ and $A(\mathbf k) = \sqrt{\abs{\bm{\lambda}(\mathbf k)}^2 + t_\perp(\mathbf k)^2}$.
It should be noted that at the level of the electronic dispersion, spin-orbit coupling enters in the same manner as interlayer coupling and it is difficult to disentangle the two.
The eigenstates of this model can be expressed as tensor products of spin and layer space states as
\begin{equation}
\begin{gathered}
    \ket{b\uparrow} = \ket{\uparrow}_\sigma \otimes \ket{b}_\tau  \\
    \ket{b\downarrow} = \ket{\downarrow}_\sigma \otimes \tau^x\ket{b}_\tau 
\end{gathered}.
\label{eq:eigenstates}
\end{equation}
The states $\ket{\uparrow}_\sigma$ and $\ket{\downarrow}_\sigma$ are defined as the spin states pointing parallel or anti-parallel to $\bm{\lambda}(\mathbf k)$, i.e. helicity states, while $\ket{b=\pm}_\tau$ are the states with layer pseudo-spin parallel or anti-parallel to $(t_\perp, 0, |\bm{\lambda}|)$. 
They can be expressed as
\begin{equation}
\begin{gathered}
    \braket{\sigma}{h}_\sigma =
    \frac{1}{\sqrt 2}
    \begin{pmatrix}
        1&
        h e^{-i\phi(\mathbf{k})}
    \end{pmatrix}^T\\
    \braket{\tau}{+}_\tau =
    \begin{pmatrix}
        w(\mathbf k)&
        z(\mathbf k)
    \end{pmatrix}^T\\
    \braket{\tau}{-}_\tau =
    \begin{pmatrix}
        - z(\mathbf k)&
        w(\mathbf k)
    \end{pmatrix}^T
    \end{gathered}
\end{equation}
where $\lambda \cos(\phi(\mathbf k)) = \bm{\lambda}(\mathbf k)\cdot \hat{\mathbf{x}}$ and 
\begin{equation}
    w(\mathbf k),\ z(\mathbf k) = \sqrt{\frac{1}{2}\left(1 \pm \frac{\lambda(\mathbf k)}{A(\mathbf k)}\right)}
\end{equation}
which implicitly defines the change of basis $\psi_{\mathbf{k}} = \check{U} c_\mathbf{k}$ to eigenstate operators.
The structure of the eigenstates leads to two Kramers degenerate Fermi surfaces, split from each other by the spin-orbit and interlayer coupling as depicted in Fig.~\ref{fig:fermi_surface}.
Each Kramers doublet consists of states with helical winding of the electronic spin about the Brillouin zone center in opposite senses.

For the purposes of calculating response functions in \cref{sec:she,sec:ee} below it is convenient to re-express the Hamiltonian in the following manner.
We define the matrices
\begin{equation}
\begin{gathered}
    \check{\Sigma}_0 = \sigma_z \tau_x,\qquad \check{\Sigma}_1 = \sigma_x \tau_y,\\ \check{\Sigma}_2 = \sigma_y \tau_y,\qquad \check{\Sigma}_3 = \sigma_z \tau_0.
    \end{gathered}
\end{equation}
These matrices along with the products $\check{\Sigma}_0\check{\Sigma}_i$ are closed under the commutators 
\begin{equation}
    \comm{\check{\Sigma}_0}{\check{\Sigma}_i} = 0,\qquad
\comm{\check{\Sigma}_i}{\check{\Sigma}_j} = 2i\epsilon_{ijk} \check{\Sigma}_k
\end{equation}
and generate the algebra $\mathfrak{so}(4)\oplus\mathfrak{u}(1)$ with the $\check{\Sigma}_i$ forming an $\mathfrak{su}(2)$ sub-algebra.
In terms of these matrices along with the the modified adjoint $\bar{c} \equiv c^\dagger \check{\Sigma}_0$, the Hamiltonian is simply
\begin{equation}
    H_0 =  \sum_{\mathbf{k}} \bar{c}_{\mathbf{k}} \pqty{\xi_\mathbf{k} \check{\Sigma}_0 + \mathbf{d}(\mathbf{k}) \cdot \bm{\check{\Sigma}}}c_{\mathbf{k}}
    \label{eq:so4h}
\end{equation}
where we have defined the vector
\begin{equation}
    \mathbf{d}(\mathbf{k}) \equiv A(\mathbf{k}) \mathbf{n}(\mathbf{k}) \equiv (\lambda \sin k_x, \lambda \sin k_y, t_\perp(\mathbf{k}))^T,
    \label{eq:paulirep}
\end{equation}
with $\abs{\mathbf{d}(\mathbf{k})}\equiv A(\mathbf{k})$ and unit vector $\mathbf{n}(\mathbf{k})$.

For all calculations in this work we use intralayer hopping strengths $t_1=1$, $t_2=-0.32$, $t_3=0.16$, interlayer hopping strength $t_\perp=0.08$, and chemical potential $\mu=-1.18$.

\section{Bogoliubov-de Gennes Hamiltonian}
\label{sec:bdg}

When writing the Bogoliubov-de Gennes Hamiltonian for the superconducting state of this model, we impose several constraints in order to match empirical details of superconductivity in this system.
The order parameter in cuprates is known to belong to the $B_{1g}$ representation ($d_{x^2-y^2}$), which we enforce by hand.
We additionally restrict pairing to be between degenerate bands; pairing between bands with different energies would either require pairing at finite momentum, which is not observed, or pairing of excitations away from the Fermi surface, which is energetically disfavored.
Finally, we impose that the system remains inversion symmetric.

With these restrictions we can now write the BdG Hamiltonian for this system as
\begin{equation}
    H_\text{BdG}= \sum_{\mathbf{k}bh}\phantom{}^{'}
    \Psi^\dagger_{\mathbf{k}hb}
    \begin{bmatrix}
        \epsilon_{b}(\mathbf k)& \Delta_b f(\mathbf k)\\
        \Delta_b^* f(\mathbf k)&-\epsilon_b(\mathbf k)
    \end{bmatrix}
    \Psi_{\mathbf{k}hb}.
    \label{eq:HBdG}
\end{equation}
Here $\Delta_b$ is the order parameter for pairing in each band and $f(\mathbf{k}) = \cos k_x - \cos k_y$ is the $d$-wave form factor. 
The Nambu spinor 
    $\Psi = \begin{pmatrix}
    \psi_{\mathbf{k}}&
        \tilde{\psi}^\dagger_\mathbf{k}
    \end{pmatrix}^T$ 
is defined in terms of the eigenstate operators $\psi_{bh}$ associated with the states in Eq.~\eqref{eq:eigenstates} and their time-reverse $\tilde{\psi} = \Theta\psi\Theta^{-1}$, where $\Theta$ is the time-reversal operator.
We do this because this system has non-trivial properties under time-reversal owing to the presence of spin-orbit coupling.
The notation $\sum'$ indicates that we sum over only half the Brillouin zone to avoid double counting states that would naturally arise in this framework. 

We have written our Nambu spinors in this form because the usual procedure of considering pairing between states of opposite spin and momentum would have unfavorable consequences in systems with inversion-symmetry breaking or spin-orbit coupling, namely the superconducting gap function would no longer transforming as a representation of the point group of the system.
The order parameter would acquire an extra phase under group operations that could not be removed by a gauge transformation, and so provides an obstruction to classifying its symmetry.
By writing the BdG Hamiltonian explicitly in terms of operators and their time-reversal, however, this problem is avoided.~\cite{sergienko_order_2004, samokhin_symmetry_2015, gong_time-reversal_2017}
Additionally, one recovers the notion of separation into singlet and triplet gaps, where now this distinction is with respect to helicity instead of spin along a fixed quantization axis.

The global inversion symmetry of this system enforces that the gap is singlet in helicity space, analogously to the case of a system without spin-orbit coupling.
Note that unlike in the case of an inversion symmetry-breaking superconductor, our quasiparticle bands remain doubly degenerate.
The difference of the gap magnitude in the two bands depends on the strength of the interaction in the respective channels, which we do not focus on here.

One can diagonalize the Hamiltonian \cref{eq:HBdG} as a sum of normal BdG Hamiltonians, and the Bogoliubov quasiparticle dispersions are found to be 
\begin{equation}
E_b(\mathbf{k}) = \sqrt{\epsilon_b(\mathbf{k})^2 + \abs{\Delta_b}^2f(\mathbf{k})^2}.
\label{eq:bdg_energy}
\end{equation}
Because of the singlet nature of the gap, Bogoliubov quasiparticles are superpositions of a quasi-electron state and a quasi-hole in the corresponding time-reversed state.
These two states will have the same spin and so as a direct consequence, the Bogoliubov quasiparticles inherit the spin-texture of the normal state bands.
This means that near the nodes, there are gapless spin-orbit-coupled excitations.

Finally, we note that the parametrization introduced in \cref{eq:paulirep} can be straightforwardly extended to the BdG Hamiltonian \cref{eq:HBdG}, allowing the response functions to be neatly expressed in terms of the vector $\mathbf{d}$ and its derivatives (for more details see \cref{sec:appendix-param}).

\section{Spin-Hall Effect}
\label{sec:she}
\begin{figure}
    \centering
    \includegraphics[width=0.8\linewidth]{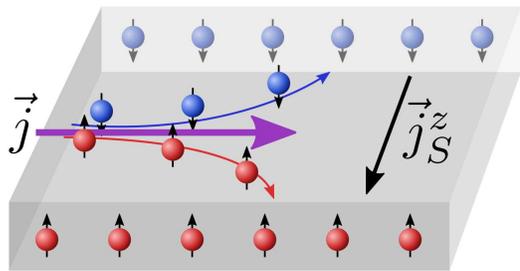}
    \caption{(Color online) A schematic depiction of the spin-Hall effect.
    The charge current $\mathbf{j}$ is split into spin-up (red) and spin-down (blue) components, which accumulate on opposing boundaries.
    This can be interpreted as a charge current inducing a transverse spin current, $\mathbf{j}_S^z$.}
    \label{fig:SHE}
\end{figure}

One of the most commonly studied spin transport effects in spin-orbit coupled materials is the spin-Hall effect (SHE), in which spin accumulates on the boundaries of a material parallel to the electrical current flowing through it, with the projection of the spin being opposite on opposing boundaries, as depicted in Fig. \ref{fig:SHE}. 
A quantity often considered in the context of the SHE is the spin-hall conductivity, which describes the flow of a spin-current perpendicular to an applied electric current, with the projection of the carried spin being perpendicular to the plane defined by the currents themselves. \cite{sinova_spin_2015}
Here we focus on the spin-hall conductivity as a hallmark of the SHE, but note that the two are not necessarily simply related, as will be discussed further below.

In particular, we are interested in the dc intrinsic spin-Hall conductivity 
\begin{equation}
    \sigma_{xy}^z = \lim_{\omega \to 0} \lim_{\mathbf{q}\to0} \frac{\Pi^\text{int}_{xy}(\mathbf{q}, i\omega_m \to \omega + i0^+)}{i\omega},
\end{equation}
written here in terms of the intrinsic contribution to the spin-Hall response,

\begin{gather}
    \Pi_{xy}^{\text{int}}(\mathbf q, \omega_m) = \left\langle j_{S,x}^z(\mathbf q, \omega_m)j_y(-\mathbf q, -\omega_m)\right\rangle, \label{eq:SHresponse}\\
    j_{S,x}^z(\mathbf q,\omega_m) = \sum_{\mathbf k,\epsilon_n} c^\dagger_{k-\frac{q}{2}} \tfrac{1}{2} \left\{v_x(\mathbf k), \sigma_z\right\} c_{k+\frac{q}{2}},\label{eq:spincurrent}\\
    j_y(\mathbf q,\omega_m) = e \sum_{\mathbf k,\epsilon_n} c^\dagger_{k-\frac{q}{2}} v_y(\mathbf k)c_{k+\frac{q}{2}}\label{eq:chargecurrent}.
\end{gather}
Here $k=(\mathbf k, \epsilon_n)$ stands for the momentum and Matsubara frequency, respectively, $e=-|e|$ is the electron charge, $v_i(\mathbf k) = \partial H_0(\mathbf k)/\partial k_i$ is the velocity operator, and $c^\dagger_k, c_k$ are the electron creation and annihilation operators.
For our analysis these operators create quasi-particles of definite spin and layer index.
Equation~\eqref{eq:spincurrent} gives a common convention for the spin-current, and the superscript $z$ indicates the polarization of the spin-current.\footnote{The presence of spin-orbit coupling means that there is no longer a natural definition of spin-current in terms of a conserved Noether charge.} 
Explicit calculation leads to
\begin{multline}
    \sigma^z_{xy} = 4 e\sum_{\mathbf{k}}\phantom{}^{'} \frac{\partial\xi}{\partial k_x} 
    \pqty{\frac{\partial \mathbf{d}}{\partial k_y} \times \mathbf{n}}\cdot \hat{\mathbf{z}}
    \\
    \times \left[\left(\frac{\epsilon_+}{E_+} +\frac{\epsilon_-}{E_-}\right)\frac{\pi_N}{E_+ - E_-}\right.\\
        + \left.\left(\frac{\epsilon_+}{E_+} -\frac{\epsilon_-}{E_-}\right)\frac{\pi_S}{E_+ + E_-}\right]
        \label{eq:spin_hall_conductivity}
\end{multline}
where we have defines the normal- and superfluid-like bubbles
\begin{equation} \label{eq:bubbles}
\pi_{N/S}(\mathbf{k}) = \frac{n(E_-(\mathbf{k})) - n(\pm E_+(\mathbf{k}))}{E_+(\mathbf{k}) \mp E_-(\mathbf{k})}
\end{equation}
with $n$ the quasiparticle occupation function, and $\mathbf{n}$ and $\mathbf{d}$ the vectors introduced in the parametrization \cref{eq:paulirep}. 
We have suppressed the dependence on the momentum $\mathbf{k}$.


Spin-orbit coupling will in general lead to a non-zero spin-Hall conductivity.
There are notable exceptions, however, where the spin-Hall conductivity exactly vanishes, and the exact conditions under which it remains finite, particularly for Rashba spin-orbit coupling, have been a subject of lively discussion.\cite{sinova_universal_2004,inoue_suppression_2004,mishchenko_spin_2004,dimitrova_vanishing_2004,dimitrova_spin-hall_2005,krotkov_intrinsic_2006,tse_spin_2006,galitski_boundary_2006,bleibaum_boundary_2006}
These arguments for a vanishing of the spin-Hall conductivity do not hold in our model, however, and we find a non-zero result. 

\begin{figure}
    \centering
    \includegraphics[width=\linewidth]{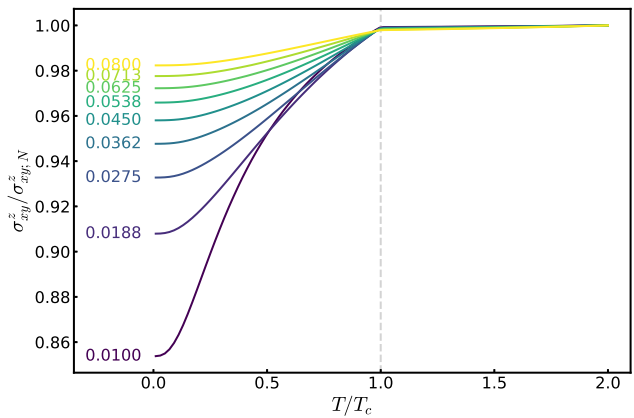}
    \caption{(Color online) The total DC spin-Hall conductivity $\sigma_{xy}^z$ as a function of temperature for different values of the spin-orbit coupling strength $\lambda$.
    The conductivity is roughly constant in the normal phase.
    We therefore normalize all results by the value of the spin-hall conductivity in the normal phase for the corresponding choice of $\lambda$.
    The spin hall conductivity however exhibits a decrease with the onset of superconductivity.
    This can be attributed to a finite energy for reorienting spins associated with the formation of singlet-like Cooper pairs in the superconducting phase.
    In these calculations, we have chosen $T_c=0.016t_1$.
    \label{fig:spinhall}}
\end{figure}

One of the main issues regarding the consideration of the spin-Hall conductivity is the difficulty in relating it to experimental measurements of the spin-Hall effect. 
Since spin is not a conserved quantity in a system with spin-orbit coupling, it does not obey a continuity equation, so a spin-current cannot be consistently and rigorously defined; Eq.~\eqref{eq:spincurrent}, which we use in our calculation, is a common choice, but is not uniquely determined.
Therefore, the accumulation of spin at the boundaries of the system is not directly related to the spin-Hall conductivity in the same way that accumulation of charge is related to electrical Hall conductivity. 

This can be most easily demonstrated by considering the transformation properties of spin and spin-current under time reversal.
As pointed out by Rashba,\cite{rashba_spin-orbit_2006} the definition of the spin current used in defining the spin-Hall conductivity is even under time reversal, while magnetization measured at sample boundaries is odd under the same operation.
Consequently, there must be some additional time-reversal-symmetry breaking process that relates spin-current to spin accumulation at sample boundaries, which is not included in calculations of the spin-Hall conductivity.
Indeed, we note that we find a \emph{total} spin-Hall conductivity, despite the fact that the system does not break global inversion symmetry.
Furthermore, the sense of the spin-Hall conductivity does not depend on the sign of the spin-orbit coupling.
This is a strong indication that the spin-Hall conductivity as traditionally calculated is not directly related to an observable.

Nonetheless, we have included this calculation of the spin-Hall conductivity for completeness.
For the aforementioned reasons it is not clear how to relate this result to a precise experimental prediction, except to note that a non-zero spin-Hall conductivity typically indicates that a spin-Hall effect can be observed.
Because the spin-orbit coupling is seen to change sense between layers, we may expect whatever spin-Hall response there is in experiment to be layer-staggered as well.

\section{Edelstein Effect}
\label{sec:ee}
Another frequently considered spintronic effect is the Edelstein effect, also called the inverse spin-galvanic effect (ISGE).
In the Edelstein effect a charge current generates a uniform spin polarization throughout the bulk of a SOC material.\cite{edelstein_spin_1990}
Traditionally the effect is discussed in the context of applying an electric field and observing the resultant spin polarization, and such behavior has been experimentally observed.\cite{silov_currentinduced_2004,kato_observation_2004}

\begin{figure}
    \centering
    \includegraphics[width=0.8\linewidth]{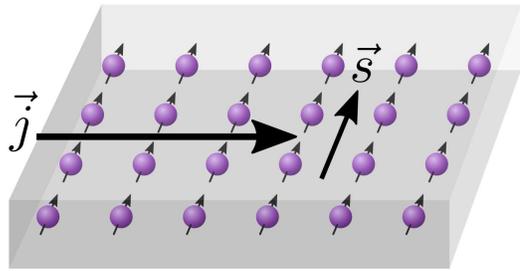}
    \caption{(Color online) A schematic depiction of the Edelstein effect.
    The charge current $\mathbf{j}$ induces a uniform transverse spin polarization $\mathbf{s} = \hat{\alpha}_\mathbf{EE} \mathbf{j}$  throughout the bulk of the material.}
    \label{fig:EdelSchem}
\end{figure}

The effect can typically be quantified through the dc Edelstein conductivity $\sigma_\text{EE}$ given by
\begin{equation}\label{eq:EEcond}
\begin{gathered}
    \chi_\text{EE}(\mathbf{q}, i\omega_m) = \ev{s_x(\mathbf{q}, i\omega_m)j_y(-\mathbf{q},-i\omega_m)}\\
   \sigma_\text{EE} =  
    \lim_{\omega \to 0} \lim_{\mathbf{q}\to0} \frac{\chi_\text{EE}(\mathbf{q}, i\omega_m\to\omega + i0^+)}{i\omega},
\end{gathered}
\end{equation}
which defines the linear response relationship $s_x=\sigma_\text{EE}E_y$.
This relation is, however, problematic in the context of a superconductor.
As stated above, the Edelstein effect is, properly, the relationship between spin polarization and current flow.
In the normal state the current is directly related to an applied electric field through the dc electrical conductivity $j_i = \sigma E_i$, and so $\sigma_\text{EE}$ is an accurate proxy for the strength of the effect. 
However, in a superconductor (or the pathological case of a metal without disorder), the electrical conductivity $\sigma$ is infinite, so application of an electric field does not result in a steady state current, and the dc Edelstein conductivity in Eq.~\eqref{eq:EEcond} is not a meaningful quantity.

Instead, in order to consider the Edelstein effect in a superconductor we need to directly relate the spin polarization and the supercurrent as $s_x=\alpha_\text{EE}j_y$, with $\alpha_\text{EE}$ now being the quantity of primary interest.
Considering the finite frequency response of the system, we can relate $\alpha_\text{EE}(\omega)$ to the Edelstein susceptibility $\chi_\text{EE}(\omega)$ above and the electromagnetic response function $\Pi(\omega)$ from which one obtains the dc electrical conductivity as $\sigma = \lim_{\omega\to 0}\Pi(\omega)/(i\omega)$.
We have
\begin{multline}
\label{eq:alpha-relation}
    s_x(\omega) = \alpha_\text{EE}(\omega) \underbrace{\frac{\Pi(\omega)}{i \omega} E_y(\omega)}_{=j_y(\omega)}
    = \frac{\chi_\text{EE}(\omega)}{i\omega}E_y(\omega)\\
    \implies \alpha_\text{EE}(\omega) = \frac{\chi_\text{EE}(\omega)}{\Pi(\omega)},
\end{multline}
giving $\alpha_\text{EE}(\omega)$, for which the dc limit $\omega\to 0$ is finite.
This result will also be useful in evaluating the normal state response since our model contains no disorder effects.

We can understand, at a heuristic level, how the current and spin can be related through the following argument.
Let us consider the case where this model contains a supercurrent.
The Cooper pairs then acquire finite momentum $\mathbf{Q}$.
We can absorb this momentum by making the gauge transformation $\mathbf{A} \to \mathbf{A} + \mathbf{Q}/(2e)$.
To lowest order in $\mathbf{Q}$ the action is shifted to
$S - j \cdot Q/(2e)$.
Computing the spin expectation value of the system, we find to lowest order in $\mathbf{Q}$
\begin{equation}
    \ev{s_x} = -\frac{Q}{2e}\ev{s_x j_y} = 
    -\frac{Q}{2e} \chi_{EE}(\mathbf{0}, \omega \to 0).
\end{equation}
From Ginzburg-Landau theory we have that the supercurrent is
\begin{equation}
   j^y = -\frac{e}{m} n_s Q, 
\end{equation}
where $n_s$ is the density of superfluid electrons, and we assume that quasiparticles do not significantly contribute to the current, so this approximately represents the entire charge current of the system.
We thus have that in the superconducting state
\begin{equation}\label{eq:naivealpha}
    \ev{s_x} = \frac{m \chi_{EE}}{2e^2 n_s} j_y.
\end{equation}
So in the case of a uniform supercurrent the non-vanishing of the Edelstein susceptibility $\chi_\text{EE}$ implies a relationship between the supercurrent and spin-polarization, as expected from \cref{eq:alpha-relation}.
Such a relation was studied in the case of an s-wave pairing previously.\cite{edelstein_magnetoelectric_1995}

We note, however, that because our model is inversion symmetric, there can be no total spin polarization in the system, $s_x = 0$.
However, the \emph{layer-staggered} analogs of \cref{eq:alpha-relation,eq:naivealpha}, corresponding to the response of the \emph{layer-staggered} spin $\tilde{s}_x \equiv s_x\tau_z$, behave in much the same manner.

\begin{figure}[!tp]
    \centering
    \includegraphics[width=\linewidth]{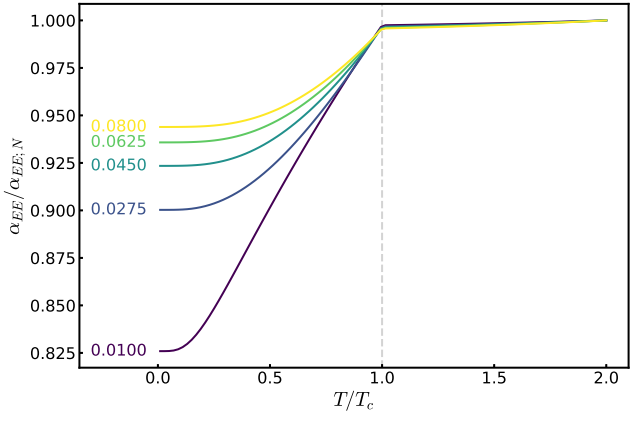}
    \caption{(Color online) The Edelstein susceptibility $\alpha_\text{EE}$ relating the layer-staggered polarization to the transverse current via $\tilde{s}_x = \alpha_\text{EE}j_y$ as depicted schematically in \cref{fig:EdelSchem}.
    The coefficient $\alpha$ is plotted for a range of spin-orbit coupling strengths $\lambda$ with each line normalized by the (roughly) constant normal state Edelstein response.
    There is a marked change in the magnitude of the effect coinciding with the onset of superconductivity.
    }
    \label{fig:edelstein_bcs}
\end{figure}

This response $\tilde{s}_x = \alpha_\text{EE}j_y^S$ can obtained be via the analog of \eqref{eq:alpha-relation}, replacing the normal spin with layer-staggered spin.
Some calculation allows us to find the general formulae
\begin{multline}
        \chi(\omega \to 0) = 8e \sum_{\mathbf{k}}\phantom{}^{'} A(\mathbf{k})\frac{\partial n_y(\mathbf{k})}{\partial k_y}\\
        \times \pqty{l(\mathbf{k})^2 \pi_N(\mathbf{k}) + p(\mathbf{k})^2 \pi_S(\mathbf{k})} 
        \label{eq:chigeneral}
\end{multline}
and
\begin{multline}
    \Pi(\omega \to 0) = 
    2 e^2\sum_{\mathbf{k}}\phantom{}^{'} \\
    \times
    \left[2A(\mathbf{k})^2\abs{\frac{\partial\mathbf{n}(\mathbf{k})}{\partial k_y}}^2
    \left( l(\mathbf{k})^2 \pi_N(\mathbf{k}) + p(\mathbf{k})^2 \pi_S(\mathbf{k})\right)\right.\\
    \left.+ \sum_b \left(\frac{\partial^2 \epsilon_b(\mathbf{k})}{\partial k_y^2}
    -b A(\mathbf{k})\abs{\frac{\partial\mathbf{n}(\mathbf{k})}{\partial k_y}}^2
    \right) \frac{\epsilon_b(\mathbf{k})}{2 E_b(\mathbf{k})}\tanh\frac{E_b(\mathbf{k})}{2T} \right],
    \label{eq:pigeneral}
\end{multline}
where we have defined the coherence functions
\begin{equation}
l(\mathbf{k})^2,p(\mathbf{k})^2 = 1 \pm \frac{\epsilon_+(\mathbf{k}) \epsilon_-(\mathbf{k}) + \Delta_+ \Delta_- f_\mathbf{k}^2}{E_+(\mathbf{k}) E_-(\mathbf{k})},
\end{equation}
$\mathbf{n}$ is again the unit vector introduced in \cref{eq:paulirep}, and $\pi_{N/S}$ are the quantities defined in \cref{eq:bubbles}.
The corresponding expressions for the normal state are found as the $\Delta\to 0$ limit of these. 
The quantity $\alpha_\text{EE}$, obtained as $\lim_{\omega\to0}\chi(\omega)/\Pi(\omega)$, is plotted in \cref{fig:edelstein_bcs} as a function of temperature across the superconducting transition.
Whereas the effect is only weakly dependent on temperature above $T_c$, in the superconducting phase the magnitude of the effect can change dramatically with temperature, especially for weaker spin-orbit coupling.

The result becomes more complicated in the case where the current is non-uniform or if the two gaps have different phase structures.
To obtain insight into this case, as well as the above linear response calculation, we now derive a Ginzburg-Landau-like generating functional for the layer-staggered spin-density.
We start with the Bogoliubov-de Gennes (BdG) action $S_\text{BdG}$ along with the associated Hubbard-Stratonovich terms.
The first modification is to give the order parameter a spatially varying phase in order to describe a supercurrent carrying state.
The $\Delta$ terms now connect states of momentum $\mathbf{k} + \mathbf{Q}/2$ and $\mathbf{k} - \mathbf{Q}/2$ where $\mathbf{Q}$ is the Cooper pair momentum.
Secondly we will add a source field for layer staggered spin-density, which takes the form $B\bar{\psi} \tilde{s}_x \psi$.
Integrating out the Fermions we obtain a generating functional for the staggered spin density $Z[\Delta, \mathbf{Q}, B]$ such that $\ev{\tilde{s}_x} = -T\frac{d}{dB}\ln Z|_{B=0}.$

Our next step is to approximate the generating functional within a Ginzburg-Landau approximation
\begin{equation}
    Z = e^{-\beta F} \approx e^{-\beta F_\text{GL}}
\end{equation}
where
\begin{multline}
    F_\text{GL} = \int d\mathbf{r}\left(
        \alpha_b |\Delta_b|^2 + K_{bb'}\mathbf{D}\Delta^*_b
        \mathbf{D}\Delta_{b'}
        \right.\\
        \left.+ \beta_b |\Delta_b|^4
        + B K^{xy}_{bb'} \Delta_b^*(-i\mathbf{D})\Delta_{b'}
    \right)
\end{multline}
and $\mathbf{D} = \nabla - 2i e\mathbf{A}$ is the covariant derivative, with sums over repeated indices.
To do so we start with the standard Hubbard-Stratonovich decoupled form of the mean-field problem, including now the layer-staggered source term
\begin{equation}
   S = \sum_b \frac{1}{g_b} |\Delta_b|^2 + S_{BdG}[\Delta_b]
   + \sum_k\phantom{}^{'} B\bar{\psi}\tilde{s}_x\psi
   .
\end{equation}
We then integrate out the gaussian fermionic theory to obtain
\begin{equation}
   S = \sum_b \frac{1}{g_b} |\Delta_b|^2 - \Tr\ln\left[\hat{G}^{-1}_0 - \Delta_b \hat{V}_{\Delta;b} - B \hat{V}_B\right].
\end{equation}
Performing a 4th order gradient expansion in $\Delta$ and keeping only to lowest order in $B$ we obtain the explicit form of the Ginzburg-Landau coefficients\footnote{We note that the coefficients obtained here differ from what one would obtain by expanding \cref{eq:chigeneral,eq:pigeneral}.
This is due to different order of limits between the two expressions. Here we take the limit $\omega \to 0$ before $\mathbf{q} \to 0$ reflecting the fact that we are considering a supercurrent carrying system in thermal equilibrium.
In contrast, the general expressions derived above consider the case where the system is responding to the presence of a supercurrent which has been ``switched on''.
The general expressions above can be derived in the opposite limit without much extra complication leading to additional intra-band terms in the response.
}
\begin{equation}
    \begin{gathered}
    \alpha_b = \frac{1}{g_b} - \sum_{\mathbf k}\phantom{}^{'} \frac{f(\mathbf k)^2 \tanh\left(\frac{\epsilon_b}{2T}\right)}{\epsilon_b}\\
    \beta_b = \sum_{\mathbf k}\phantom{}^{'} \frac{f(\mathbf k)^4}{2 \epsilon_b^2}c(\epsilon_b)\\
    \begin{split}
    K^{xy}_{bb}= b\sum_{\mathbf k}\phantom{}^{'} \frac{f(\mathbf k)^2}{\epsilon_b} \left[n_y\frac{\partial \epsilon_b}{\partial k_y} \pqty{n''(\epsilon_b) + \frac{c(\epsilon_b)}{\epsilon_b}}\right.\\
    \left.+ \frac{\partial n_y}{\partial k_y} \pqty{c(\epsilon_b)
    + \epsilon_+ \epsilon_-S(\epsilon_+, \epsilon_-)
    }\right]
    \end{split}\\
    K^{xy}_{b,-b} = -2\sum_{\mathbf k}\phantom{}^{'} f(\mathbf k)^2 
    A \frac{\partial n_y}{\partial k_y} S(\epsilon_+, \epsilon_-)\\
    \begin{split}
    K_{bb} =
    \sum_{\mathbf k}\phantom{}^{'}
        \frac{f(\mathbf k)^2}{4 \epsilon_b}
        \left[
            \frac{\partial^2 \epsilon_b}{\partial k_x^2} c(\epsilon_b)
            + \left(\frac{\partial \epsilon_b}{\partial k_x}\right)^2 n''(\epsilon_b)\right.\\
            \left.
            + 2 A^2 \left\vert\frac{\partial \mathbf{n}}{\partial k_y}\right\vert^2 \frac{\epsilon_+ \epsilon_-}{\epsilon_b - \epsilon_{-b}} S(\epsilon_+ \epsilon_-) \right]
            \end{split}
            \\
    K_{b,-b} =
   -\frac{1}{4} \sum_{\mathbf k}\phantom{}^{'}
        f(\mathbf k)^2A^2 \left\vert\frac{\partial \mathbf{n}}{\partial k_y}\right\vert^2 S(\epsilon_+, \epsilon_-)
    \end{gathered}
\end{equation}
where
\begin{equation}
\begin{gathered}
        c(x) = \frac{\tanh\left(\frac{x}{2T}\right)}{2x} + n'(x),\\
   S(x, y) = \frac{\frac{1}{x}\tanh{\frac{x}{2T}} - \frac{1}{y}\tanh{\frac{y}{2T}}}{x^2-y^2},
\end{gathered}
\end{equation}
$n$ is the Fermi function, and we recall that $\sum'$ indicates summation over half the Brillouin zone.
The interesting magneto-electro effects are due to the the presence of the $K^{xy}$ terms, sometimes called Lifshitz invariants, allowed by the breaking of inversion symmetry within each layer\cite{mineev_nonuniform_2008}, which arise from the $\Tr\left[(\hat{G}\hat{V}_\Delta)^3\hat{G}\hat{V}_B\right]$ term of the above expansion.
In our case, instead of coupling to the magnetic field, these terms couple to the generating field of layer-staggered spin.


With this we can now see how the Edelstein effect arises from Ginzburg-Landau theory. 
Suppose we have a uniform supercurrent in the $y$ direction. 
From the Ginzburg-Landau theory, we have that
\begin{equation}
    j_y = 2 e \sum_{bb'} K_{bb'} \Delta_b^* \Delta_{b'} Q_y.
\end{equation}
Recall that since $F$ is a generating functional, we also have
\begin{equation}
    \ev{\tilde{s}_x}= \sum_{bb'} K^{xy}_{bb'} \Delta_b^* \Delta_{b'} Q_y,
\end{equation}
and we can then write
\begin{equation}
    \ev{\tilde{s}_x}= \frac{\sum_{bb'} K^{xy}_{bb'} \Delta_b^* \Delta_{b'}}{2 e \sum_{bb'} K_{bb'} \Delta_b^*\Delta_{b'}}
    j_y,
\end{equation}
giving the Ginzburg-Landau theory equivalent of the quantity $\alpha_\text{EE}$ calculated above from linear response. 
In the more general case, the expression becomes
\begin{equation}
    \ev{\tilde{s}_x}= \frac{\sum_{bb'} K^{xy}_{bb'} \Delta_b^* \partial_y \Delta_{b'}}{2 e \sum_{bb'} K_{bb'} \Delta_b^*\partial_y\Delta_{b'}}
    j_y,
\end{equation}
Similar effects have been derived from the Ginzburg-Landau free energies of inversion-symmetry breaking superconductors\cite{edelstein_magnetoelectric_1995,konschelle_theory_2015}, but the layer-staggered polarization predicted here is novel in the cuprates.

One might expect from the above that since an analog of the ISGE exists in bilayer cuprates, so might an analog of the spin Galvanic effect, which would allow a supercurrent to be driven by application of a static magnetic field.
In inversion asymmetric systems such behavior is generally preempted by the transition to a helical superconducting phase with zero net supercurrent.\cite{agterberg_novel_2003,dimitrova_theory_2007,bauer_magnetoelectric_2012,smidman_superconductivity_2017}
It remains to be investigated whether similar reasoning rules out the spin Galvanic effect in this system. 

\section{Discussion and Conclusions}
\label{sec:end}

In this work we have considered a model of spin-orbit coupling the superconducting state of a bilayer cuprate superconductor.
We have shown that the inversion-symmetry-preserving spin-orbit coupling posited to be present in Bi2212\cite{Gotlieb2018} should lead to non-trivial layer polarized spin-orbit effects.

In particular, we calculate the layer polarized spin-Hall conductivity, which was found to be non-trivial.
While this quantity cannot be directly related to a measurable quantity, such as the accumulation of spin at sample boundaries, we nonetheless expect that a layer-staggered spin-Hall effect should be present.

More interestingly, the presence of the new coupling term leads to a layer-staggered analog of the Edelstein (or inverse spin-galvanic) effect, an in-plane spin polarization in the presence of an applied supercurrent. 
This should be visible through optical methods such as Faraday rotation\cite{kato_currentinduced_2004,kato_observation_2004} or by measuring the degree of circular polarization in photoluminescence.\cite{silov_currentinduced_2004}
Furthermore, one could attempt ARPES measurements in a supercurrent-carrying state to directly see that canting of the in-plane spins due to this effect.\cite{Boyd2014}

There are still interesting effects to consider beyond what we have looked at in this work.
In particular, spin-resolved ARPES observes a non-trivial variation of spin texture within the Brillouin zone, most notably including a reversal of the spin texture as a function from the zone center.\cite{Gotlieb2018}
This suggests a more complicated form of the spin orbit coupling which could lead to further effects.
Additionally, for $d_{x^2-y^2}$ superconductivity a gradient in the $d$-wave order parameter can admix an $s$-wave pairing term through a coupling of the gradients.~\cite{li_mixed_1993,berlinsky_ginzburg-landau_1995,feder_microscopic_1997}
Regardless, the presence of spin-orbit coupling in BSCCO should lead to the presence of a multitude of fascinating spin-orbit driven effects, including the spin-Hall effect, Edelstein effect, and more.

\begin{acknowledgments}
We would like to thank A. Lanzara, K. Gotlieb, C.-Y Lin, M. Serbyn, I. Appelbaum, V. Yakovenko, and V. Stanev for enlightening discussions.
This work was supported by DOE-BES (DESC0001911) and the Simons Foundation (V.G. and A.A.) and NSF DMR-1613029 (Z.R.).
\end{acknowledgments}

\appendix
\section{Parametrization of the BdG Hamiltonian}
\label{sec:appendix-param}

For purposes of calculation, it is convenient to parametrize the BdG Hamiltonian, and associated Nambu Green's function in terms of the vector $\mathbf{d}$ and $\check{\Sigma}$ matrices introduced in \cref{eq:paulirep}. To do so we note that the matrices $\check{P}_\pm \equiv \pqty{\check{\Sigma}_0 \pm \mathbf{n} \cdot \check{\mathbf{\Sigma}}}/2$ satisfy the the relation
\begin{equation}
    \check{P}_b \mathbf{n} \cdot \check{\mathbf{\Sigma}} \check{P}_{b'}
    = b\delta_{bb'}\check{P}_b,
\end{equation}
meaning that they act as projectors on the degenerate eigenspaces of the normal state Hamiltonian, with the somewhat non-standard properties
\begin{equation}
    \begin{gathered}
    \check{P}_b \check{P}_{b'} = \check{\Sigma}_0 \check{P}_b \delta_{bb'},\\
    \mathbf{n}\cdot\check{\mathbf{\Sigma}} \check{P}_b = b\check{\Sigma}_0 \check{P}_b.
    \end{gathered}
\end{equation}
Using these matrices we can compactly express the BdG Hamiltonian
\begin{equation}
    H_\text{BdG} = \sum_{\mathbf{k}}\phantom{}^{'} \bar{\Psi}_{\mathbf{k}}
    \sum_b
    \bqty{
\epsilon_b(\mathbf{k})\rho_3 + \Delta_b f(\mathbf{k}) \rho_1
    }\check{P}_b
    \Psi_{\mathbf{k}}
\end{equation}
where we have introduced the adjoint Nambu spinor $\bar{\Psi} = \Psi^\dagger \check{\Sigma}_0$ and the $\rho_i$ are Pauli matrices in Nambu space.
The corresponding Nambu Green's function is found, completely analogous to the usual expresion, to be
\begin{equation}
\check{G}(i \epsilon_n, \mathbf{k}) = \sum_b \check{P}_b
\frac{i\epsilon_n\rho_0 + \epsilon_b(\mathbf{k})\rho_3 + \Delta_b f(\mathbf{k})\rho_1}{(i\epsilon_n)^2 - E_b^2}
\end{equation}
with the energies $E_b$ as defined in \cref{eq:bdg_energy}.

\section{Spin Susceptibility and Knight Shift}
\begin{figure}
    \centering
    \includegraphics[width=\linewidth]{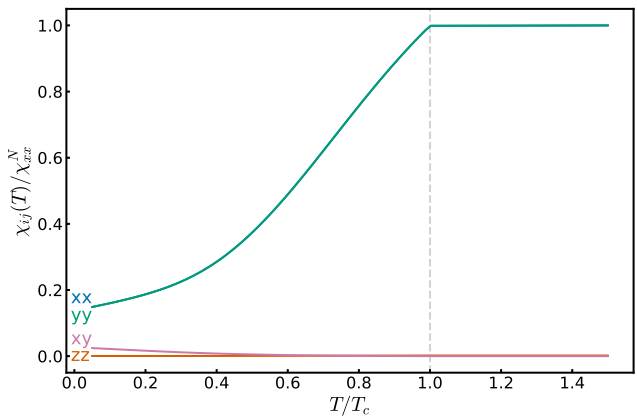}
    \caption{(Color online) The components of the static spin susceptiblity $\chi_{ij}(0, \mathbf{0})$ as a function of temperature for the layer spin-orbit coupled model.
    The diagonal in-plane components have a noticeable decrease below $T_c$ but do not go exactly to $0$ at zero temperature.
    This behavior is consistent with the decrease normally attributed to spin-singlet superconductivity in the cuprates.
    \label{fig:spinsusc}}
\end{figure}

As the presence of spin-orbit coupling induces a triplet component to the Gor'kov anomalous Green's function one might wonder why this is not in general seen in experimental signatures such as the Knight shift, where the observed behavior is seen to be consistent with singlet pairing.\cite{takigawa_spin_1989,barrett_63_1990}
In general, the telltale sign of triplet pairing in the Knight shift is that the spin-susceptibility remains constant across the superconducting transition.
On the other hand, for the case of singlet pairing there is a noticeable decrease.\cite{yosida_paramagnetic_1958,schrieffer_theory_1999}
This is, however, consistent with this model as the singlet component of the order parameter still leads to qualitative behavior similar to the typical singlet case i.e. the Knight shift should rapidly decrease below $T_c$.
However, unlike the pure singlet case as can be seen in Fig.~\ref{fig:spinsusc}, the spin-susceptibility does not go exactly to zero at $T=0$, a fact which was already appreciated by Gor'kov and Rashba in 2001.\cite{gorkov_superconducting_2001}

\bibliography{references}

\end{document}